%% file: main.tex
\documentclass{article}
\usepackage{arxiv}
\input{initializations}

\title{Transfer Learning with convolutional networks for Atmospheric parameter retrieval}

\author{
  David Malmgren-Hansen \\
  Department of Applied Mathematics and Computer Science \\
  Technical University of Denmark \\
  Denmark\\
  \And
  Allan Aasbjerg Nielsen \\
  Department of Applied Mathematics and Computer Science \\
  Technical University of Denmark \\
  Denmark\\
  \And
  Valero Laparra \\
  Image Processing Laboratory \\
  Universitat de Val{\`e}ncia\\
  Val{\`e}ncia, Spain\\
  \texttt{valero.laparra@uv.es} \\
  \And
  Gustau Camps-Valls \\
  Image Processing Laboratory \\
  Universitat de Val{\`e}ncia\\
  Val{\`e}ncia, Spain\\
  \texttt{gcamps@uv.es} \\
}

\begin{document}

\begin{center}
    ©IEEE. ACCEPTED FOR PUBLICATION IN IEEE IGARSS 2018. DOI 10.1109/IGARSS.2018.8518097\footnote{
©IEEE. Personal use of this material is permitted.  Permission from IEEE must be obtained for all other users,including reprinting/republishing this material for advertising or promotional purposes, creating new collective works for resale or redistribution to servers or lists, or reuse of any copyrighted components of this work in other works.  DOI: 10.1109/IGARSS.2018.8518097.
}
\end{center}

\maketitle

\input{sections/abstract}
\input{sections/keywords}
\input{sections/introduction}

\input{sections/method}

\input{sections/experiments}

\input{sections/conclusion}
\input{sections/references}

\end{document}

%% file: initializations.tex
\usepackage{epsfig}
\usepackage{etex}
\usepackage{amsfonts}
\usepackage{framed}
\usepackage{amssymb}
\usepackage{amsbsy} 
\usepackage{amsmath}
\usepackage{caption}
\usepackage{subcaption}
\usepackage[T1]{fontenc}
\usepackage[latin1]{inputenc}
\usepackage{multirow}
\usepackage{amsmath,epsfig}
\usepackage{array}
\usepackage[normalem]{ulem} 
\usepackage{times}
\usepackage[all]{xy}
\usepackage{flushend}
\usepackage{url}
\usepackage{rotating}
\usepackage{epstopdf}
\usepackage{float}
\usepackage[usenames, dvipsnames]{color}
\usepackage{lscape}
\usepackage{cite}
\usepackage{balance}
\usepackage{hyperref}
\hypersetup{
    bookmarks=true,         
    unicode=false,          
    pdftoolbar=true,        
    pdfmenubar=true,        
    pdffitwindow=false,     
    pdfstartview={FitH},    
    pdftitle={IASI deep retrievals},    
    pdfauthor={Camps-Valls},     
    pdfsubject={IASI deep retrievals},   
    pdfcreator={Camps-Valls},   
    pdfproducer={Camps-Valls}, 
    pdfkeywords={keyword1} {key2} {key3}, 
    pdfnewwindow=true,      
    colorlinks=true,       
    linkcolor=blue,          
    citecolor=blue,        
    filecolor=blue,      
    urlcolor=blue           
}

\usepackage{color}

%% file: sections/abstract.tex
\begin{abstract}
The Infrared Atmospheric Sounding Interferometer (IASI) on board the MetOp satellite series provides important measurements for Numerical Weather Prediction (NWP). Retrieving accurate atmospheric parameters from the raw data provided by IASI is a large challenge, but necessary in order to use the data in NWP models. Statistical models performance is compromised because of the extremely high spectral dimensionality and the high number of variables to be predicted simultaneously across the atmospheric column. All this poses a challenge for selecting and studying optimal models and processing schemes. 
Earlier work has shown non-linear models such as kernel methods and neural networks perform well on this task, but both schemes are computationally heavy on large quantities of data. Kernel methods do not scale well with the number of training data, and neural networks require setting critical hyperparameters. In this work we follow an alternative pathway: we study {\em transfer learning} in convolutional neural nets (CNNs) to alleviate the retraining cost by departing from proxy solutions (either features or networks) obtained from previously trained models for related variables.
We show how features extracted from the IASI data by a CNN trained to predict a physical variable can be used as inputs to another statistical method designed to predict a different physical variable at low altitude. In addition, the learned parameters can be transferred to another CNN model and obtain results equivalent to those obtained when using a CNN trained from scratch requiring only fine tuning.
\end{abstract}

%% file: sections/keywords.tex
\begin{keywords}
    Transfer Learning, Convolutional Neural networks, Infrared measurements, parameter retrieval
\end{keywords}

%% file: sections/introduction.tex
\section{Introduction}
\label{section:introduction}

Predicting atmospheric variables from satellite measurements is a key point in Earth observation. While Numerical Weather Prediction (NWP) methods are based on well tested physical models, sometimes the computational load makes their use prohibitive. Statistically-based methods can help either to substitute the NWP methods in some applications, or to provide predictions that can be used as a first guess for the NWP methods. 

Atmospheric profiles of physical variables are important for weather forecasting and atmospheric chemistry studies. However predicting them is one of the most challenging tasks due to their high dimensionality. On the other hand measurements from high spectral resolution instruments, like the Infrared Atmospheric Sounding Interferometer (IASI) sensor implemented on the MetOp satellite series, provide useful information for the estimation of these profiles~\cite{EUMETSAT-IASI-L1,TOURNIER2002}. Designing efficient statistically-based algorithms for this task is challenging not only because the high dimensionality of the input but also the high dimensionality of the output space. Both show high correlation and structure that should be exploited. 

Previous approaches based on kernel methods and neural networks actually exploited either spectral or vertical correlations~\cite{Blackwell2005,CAMPS2012,CampsValls16grsm,Laparra17}, but spatial redundancy was not explicitly included in the models. In this way, convolutional neural networks (CNNs) offer a good alternative to exploit relations in all three dimensions jointly~\cite{2018_Malmgren}. Neural networks offer an additional advantage to our multivariate regression problem: models are intrinsically multi-output and account for the cross-relations between the state vector at different altitudes. This allows us to attain smoothness, and hence consistency, across the atmospheric column in a very straightforward way. 

Nevertheless, estimating multiple physical variables simultaneously is complicated. Multiple models for different variables are necessary and they are heavy to train. In this sense, tools from {\em transfer learning} could be very useful in this setting, which we evaluate in this work. Here we investigate the use of two simple strategies of transfer learning which have been very successful in computer vision problems \cite{sharif2014cnn}. The basic idea is to reduce the cost of retraining neural nets by departing from proxy solutions obtained from previously trained models for related variables. 

The remainder of the paper is organized as follows. In \S\ref{sec:method} we introduce the methodology conducted in the experimental section. Section \S\ref{sec:results} gives empirical evidence in terms of accuracy, bias and smoothness of the estimates. Conclusions and future developments are given in \S\ref{sec:conclude}.

%% file: sections/method.tex
\begin{figure*}[t!]
\centering
\includegraphics[width=0.9\linewidth]{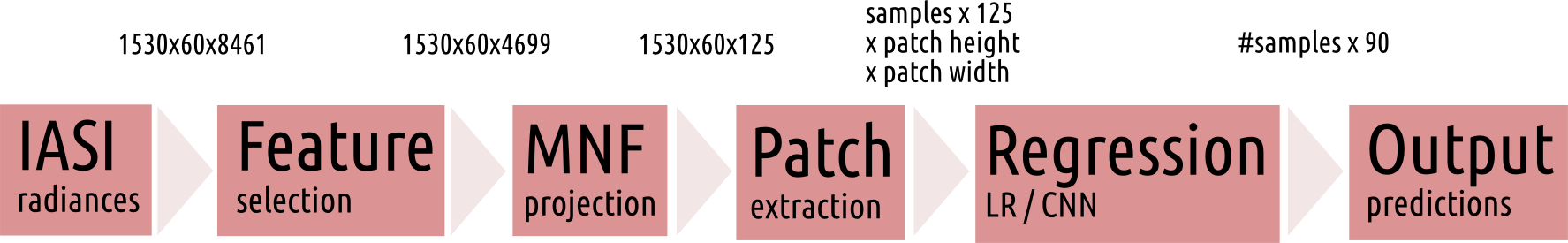}
\caption{Pipeline schematic: first selecting 4699 channels from the original 8461 spectral according to noise specifications, then performing a {\em spectral} dimensionality reduction based on linear projection to 125 features, and then a nonlinear regression based on statistically-based algorithms. Here we use linear regression (LR) and convolutional neural networks (CNN) for retrieval of atmospheric parameters at 90 altitudes simultaneously.}\label{fig:pipeline}
\end{figure*}   

\section{Methodology}\label{sec:method}


The experiments conducted here are based on the L2 processing presented in~\cite{CAMPS2012}. This scheme was proposed to predict physical atmospheric parameter profiles from the IASI data using statistically-based algorithms. We used data collected in $13$ consecutive orbits within the same day, 17-08-2013. Each orbit consists of approximately $92,000$ samples. We use the first $7$ orbits to train and the latter $6$ for testing. Note that $6-7$ orbits cover a great part of the Earth and the radiance signal in our dataset includes variances from different geographical locations.

In \cite{mnf_igarss2017} it was shown how increasing the number of spatial neighboring pixels largely improves the prediction performance when included as additional input features to standard regression models. In \cite{2018_Malmgren} we analyzed different Ordinary Least Squares Linear Regression (OLS) and CNN configurations to predict atmospheric temperature profiles. 
Table \ref{tb:rmse} shows the summary of results when using different sizes of spatial neighborhoods. It can be seen that the non-linear properties of the CNN further improves performance. 
\begin{table}[h!]
\centering
\caption{Summary of the averaged RMSE [K] across the atmosphere profile for temperature prediction.} 
\label{tb:rmse}
\vspace{2mm}
\setlength{\tabcolsep}{2pt}
\begin{tabular}{l|l|l|l|l|l|l|l|}
\hline
\multicolumn{1}{|l|}{Patch Size}     & 1$\times$1 & 3$\times$3 & 5$\times$5 & 7$\times$7 & 10$\times$10 & 15$\times$15 & 25$\times$25 \\ \hline \hline
\multicolumn{1}{|l|}{CNN}     &  --   &  2.48  & -- & -- &  2.43  & 2.20 & 2.28       \\
\multicolumn{1}{|l|}{OLS} &  3.30  &  3.00  &  2.91   &  2.86   &   2.84    &    2.85   &  --   \\ \hline
\end{tabular}
\end{table}

%% file: sections/experiments.tex
\section{Experimental results}\label{sec:results}
Despite the benefits of using CNNs for the physical variable prediction, the training process is rather expensive. The concept of transfer learning within deep learning has proven useful to overcome this problem for a range of computer vision tasks \cite{sharif2014cnn}. For example, deep CNNs trained on large databases of natural images can be transferred to smaller datasets for specific applications with high end performance. In this work, we analyze two simple yet useful strategies for transfer learning. One strategy uses a CNN trained to predict a particular physical variable as a feature extraction method. The other strategy uses the CNN parameters of the trained network as initial parameters when training a CNN to predict a different physical variable.    

\subsection{Feature extraction}

Here we explore the ability of exploiting the most successful model from previous section that was trained to predict temperature profiles to design a method to predict dew point temperature (DT) profiles. 
The straight forward way to do this uses the CNN trained to predict temperatures (\emph{CNN-T}) as a feature extraction algorithm and train a linear model to predict DT using these features as inputs. While extremely simple, this strategy has become very popular in computer vision due to its good performance \cite{sharif2014cnn}.

Note that the last layer of the \emph{CNN-T} model is a fully connected layer without rectifier, i.e. a linear combination of the outputs from the previous layer. Here we use the outputs of the second last layer (just before the last linear combination) of \emph{CNN-T} as a feature extractor, and put a linear predictor on top of it. We use this new model to predict DT, and name it  \emph{CNN-DT-F(T)}.  

\begin{figure}[t!]
\centering
\includegraphics[width=7cm]{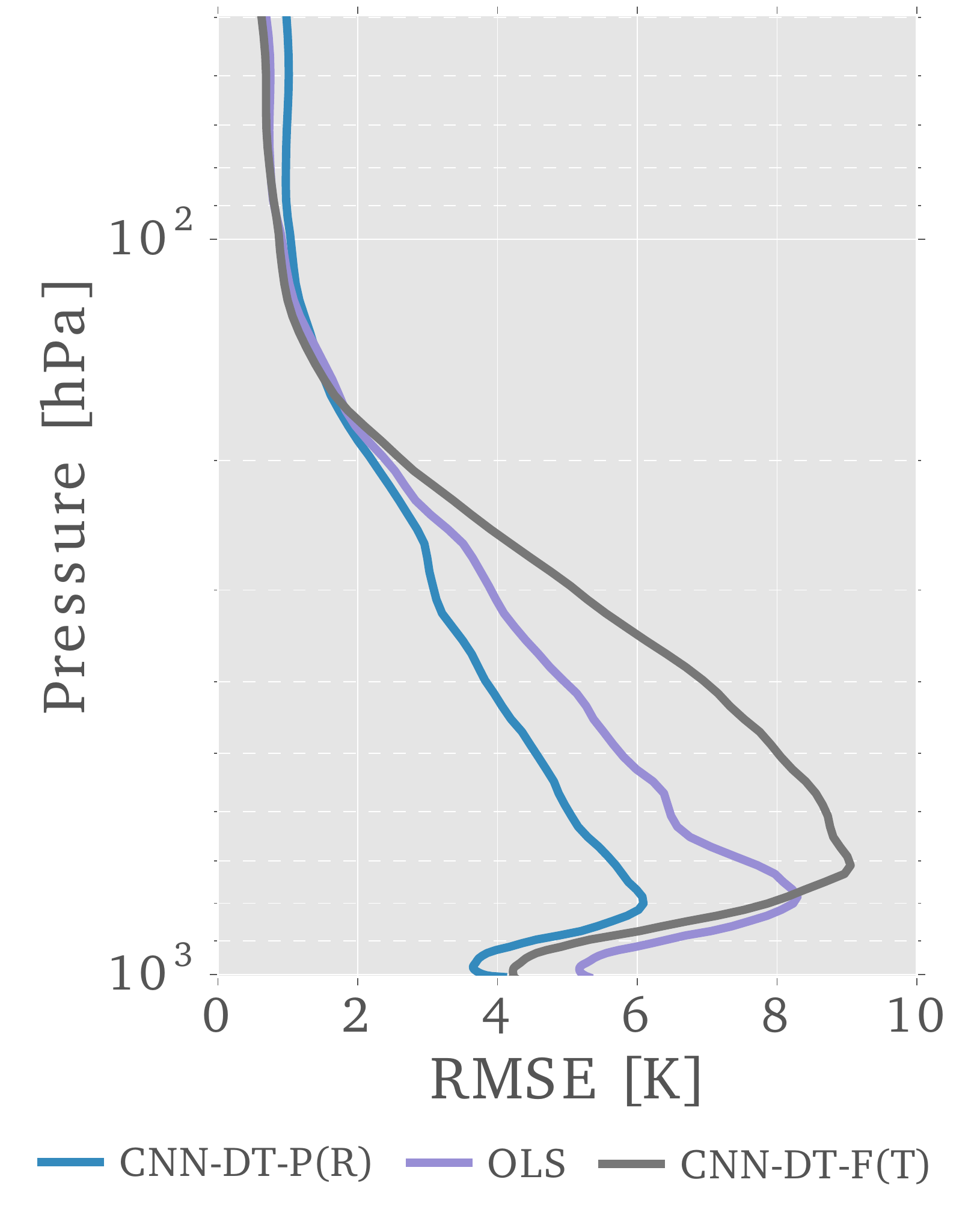}
\caption{Feature extraction approach. Dew point temperature RMSE [K] profiles for different models.}
\label{fig:transflearn1}
\end{figure}

Results obtained with such approach are presented in Fig.~\ref{fig:transflearn1}. It is interesting to compare the proposed strategy, \emph{CNN-DT-F(T)}, with a linear regression model (\emph{OLS}) since both present the same training complexity. While the proposed strategy has been previously used successfully in several works in the computer vision community, it is clear that in this case it is not efficient for all the pressure levels. In particular between [300-700] hPa the simple \emph{OLS} outperforms its performance. However for high pressure levels, the \emph{CNN-DT-F(T)} model obtains better performance than \emph{OLS}, and achieves almost the same accuracy as a CNN trained from scratch \emph{CNN-DT}. 

\subsection{Parameter initialization}

\begin{figure}[t!]
\centering
\includegraphics[width=6cm]{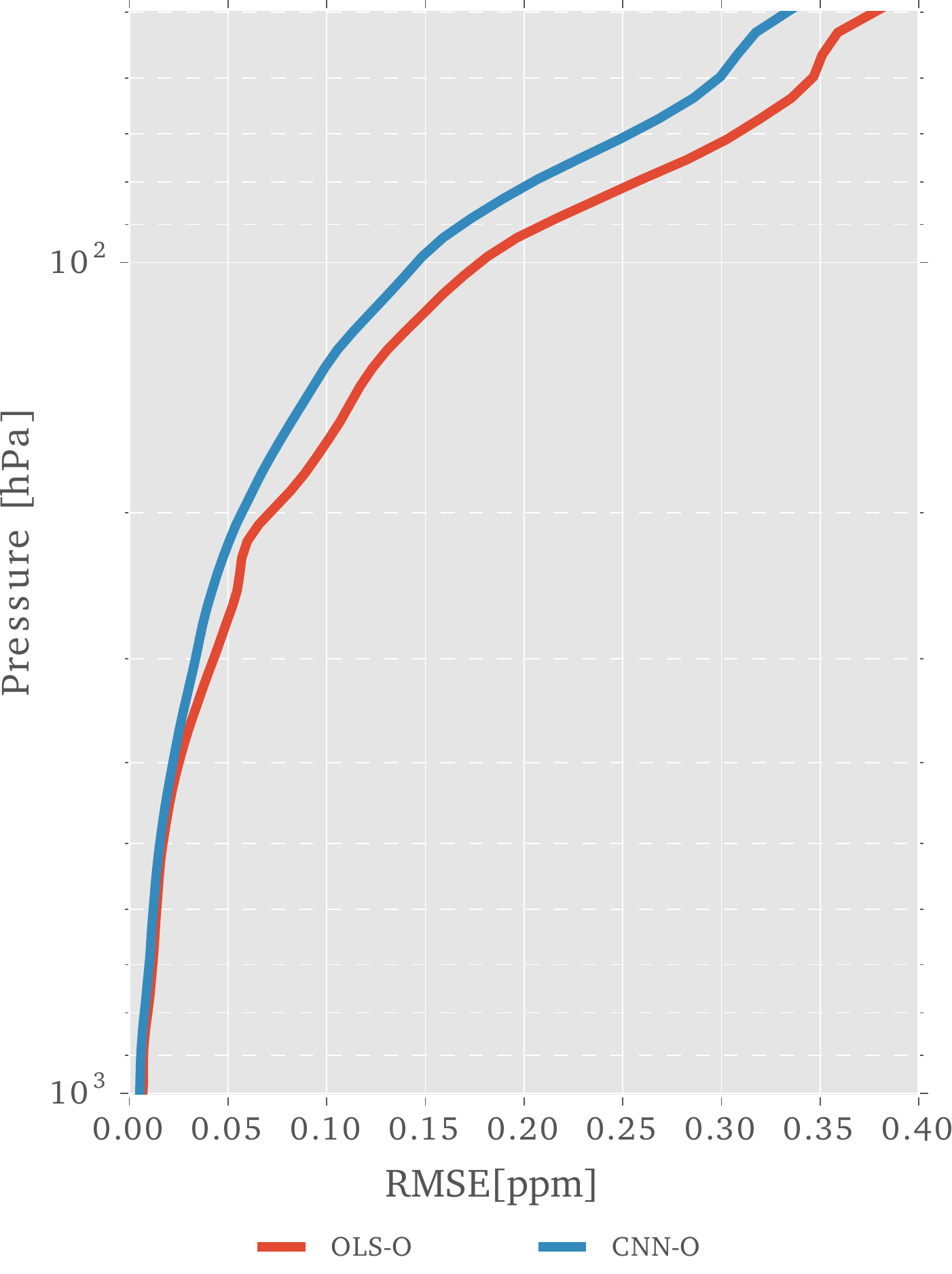}
\caption{Ozone RMSE [K] profiles of CNN and OLS regression models.}\label{fig:ozone}
\end{figure}

Although extremely simple, the previous approach has the problem of not having the warranty of obtaining an performance similar to the one obtained when using a CNN trained from scratch. Here we analyze a different strategy to use the information contained in the trained models when training new ones. We used the same network architecture as for predicting temperatures to design CNN models that predict DT. We analyze the effect of initializing the parameters for the training procedure with the ones of \emph{CNN-T}, and using the ones from a new CNN trained to predict ozone concentration [ppm] \emph{{CNN-O}}. Results for the \emph{CNN-O} network compared with a linear regression model can be seen in Fig.~\ref{fig:ozone}. This network was trained from scratch and obtained a good performance to predict ozone: RMSE [ppm] is smaller than the OLS and in the range to the ones presented in \cite{CAMPS2012} with little extra effort.

In order to compare the results, we trained a CNN from scratch using randomly initialized parameters. Therefore, we trained three new CNN models, all of them optimized to predict DT but with different parameters initialization: (1) initializing the parameters randomly, \emph{{CNN-DP-P(R)}}, (2) initializing with the \emph{CNN-T} model parameters, \emph{{CNN-DP-P(T)}}, and (3) initializing with the \emph{CNN-O} model parameters, \emph{{CNN-DP-P(O)}}. 

\begin{figure}[t!]
\centering
\includegraphics[width=7cm]{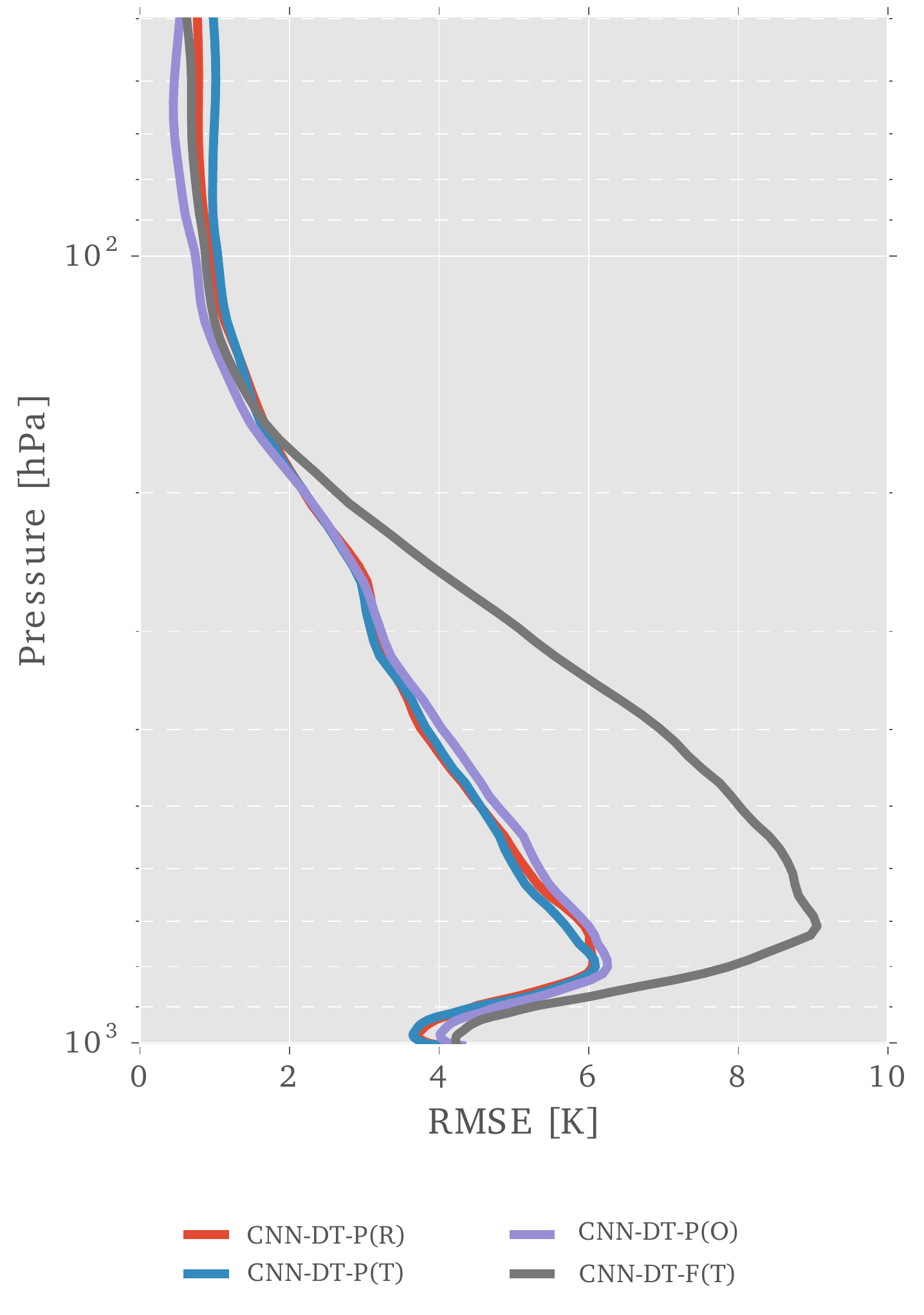}
\caption{Parameter initialization approach: RMSE profiles for DT of different regression models based on transfer learning.}\label{fig:transflearn2}
\end{figure}

Figure~\ref{fig:transflearn2} shows the results for the different networks. Several conclusions can be extracted. On the one hand, results are much better with this strategy than with the strategy used in the previous section. It is clear how the \emph{CNN-DT-P(T)} and \emph{CNN-DT-P(O)} models achieve similar performance than the \emph{CNN-DT-P(R)} model. This gives us the idea that, although the parameter space is huge, either the solutions reach the same global minimum, or the different local minima with similar performance. Both options are equivalent for practical purposes though. This is particularly interesting in the case of ozone concentration, since it has a very different nature than DT.      

\begin{figure}[t!]
\centering
\includegraphics[width=8cm]{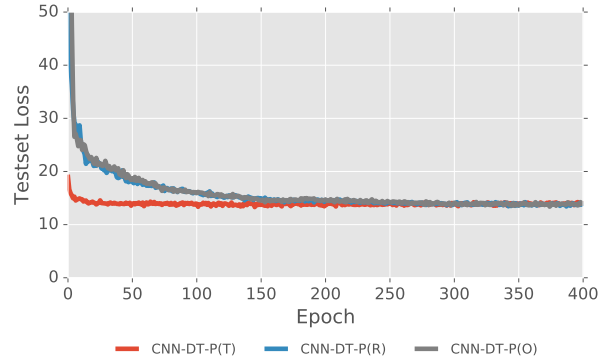}
\caption{Objective minimization during the training procedure. Performance of CNN models trained to predict DT initialized using different sets of parameters.}
\label{fig:transflearn3}
\end{figure}

Figure~\ref{fig:transflearn3} shows specifically how the objective function is minimized during the training procedure and should be used to see the convergence of each model. Regard the convergence speed, the \emph{CNN-DT-P(T)} model is much faster than the other two. However, note that \emph{CNN-DT-P(O)} converges to its minimum in a similar way as when initializing the parameters randomly \emph{CNN-DT-P(R)}. 

Although the predictions from the three different models show a similar performance in RMSE terms, solutions can fall into completely different local minimum. An interesting property of the model is the consistency of the predictions with regard other physical variables. Figure~\ref{fig:transflearn4} shows the cross-correlation matrices between the temperature predictions and the DT predictions obtained using either the \emph{CNN-DT-P(R)}, or the \emph{CNN-DT-P(T)} model. The DT predictions of the model \emph{CNN-DT-P(T)} model are similarly aligned with the temperature predictions (the cross-covariance matrix symmetry is similar) than the ones from the \emph{CNN-DT-P(R)} model. The non-diagonality (ND) has been computed as the Frobenius norm of the matrix minus its transpose, i.e. ND=$\|{\bf A}-{\bf A}^\top\|_F^2$. Surprisingly, the mutual information between the variables is smaller for the \emph{CNN-DT-P(T)} model. This is a safety check to ensure that, even that the model has been initialized with the parameters of a model trained to predict other physical variable, the predictions do not need to be dependent of the behavior of this model. Mutual information has been computed in a similar way as in \cite{mnf_igarss2017} using the method introduced in \cite{Laparra2011}.

%% file: sections/conclusion.tex
\section{Conclusions}\label{sec:conclude}

We analyzed the problem of transfer learning in multi-output physical parameter retrieval when using CNN models. In general we found that some benefits can be obtained from the transfer learning methodology. We analyzed two different strategies. Firstly, we used a model trained for predicting temperature profiles as a feature extraction method for the previous stage to a simple linear regression algorithm. We found that this strategy is not ideal but it can be helpful in some aspects. At low altitudes, we get higher accuracy than the shallow linear regression model. At higher altitudes though, the transfer learning approach does it worse. Fine tuning for a specific output variable is necessary in order to achieve good predictions. 

The second strategy was using the parameters of an already trained model as initial parameters to train a new model to predict a different physical variable, in our case dew point temperature (DT). We initialized the model using the parameters of two already trained models: one trained to predict temperature and one to predict ozone. We compared the performance also with a model trained when using randomly initialized parameters. We found that initializing the parameters using an already trained model helps in time convergence terms and achieves a similar result as training the model from scratch when the model used as initialization was trained to predict a similar variable. The performance reached by CNN initialized from random weights can be reached in less than around $\frac{1}{8}$ of the training time if the weights are transferred (initialized) from the models trained for temperature, while the one initialized with the parameters of the ozone model provides no advantage with regard the random initialization. Note that features of T and DT are more similar in structure and even in units with regard to ozone.

\begin{figure}[t!]
\centering
\begin{tabular}{cc}
ND: 19.73, MI: 0.25 & ND: 19.72, MI: 0.32
\\
\includegraphics[width=3.25cm]{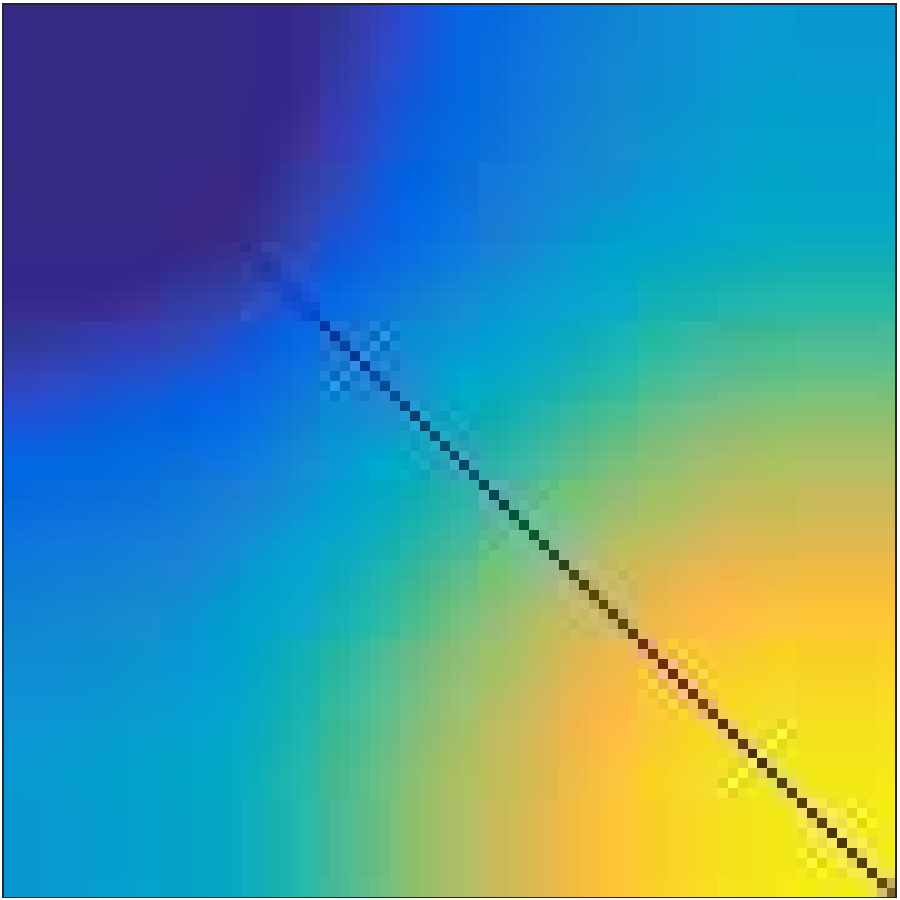} &
\includegraphics[width=3.25cm]{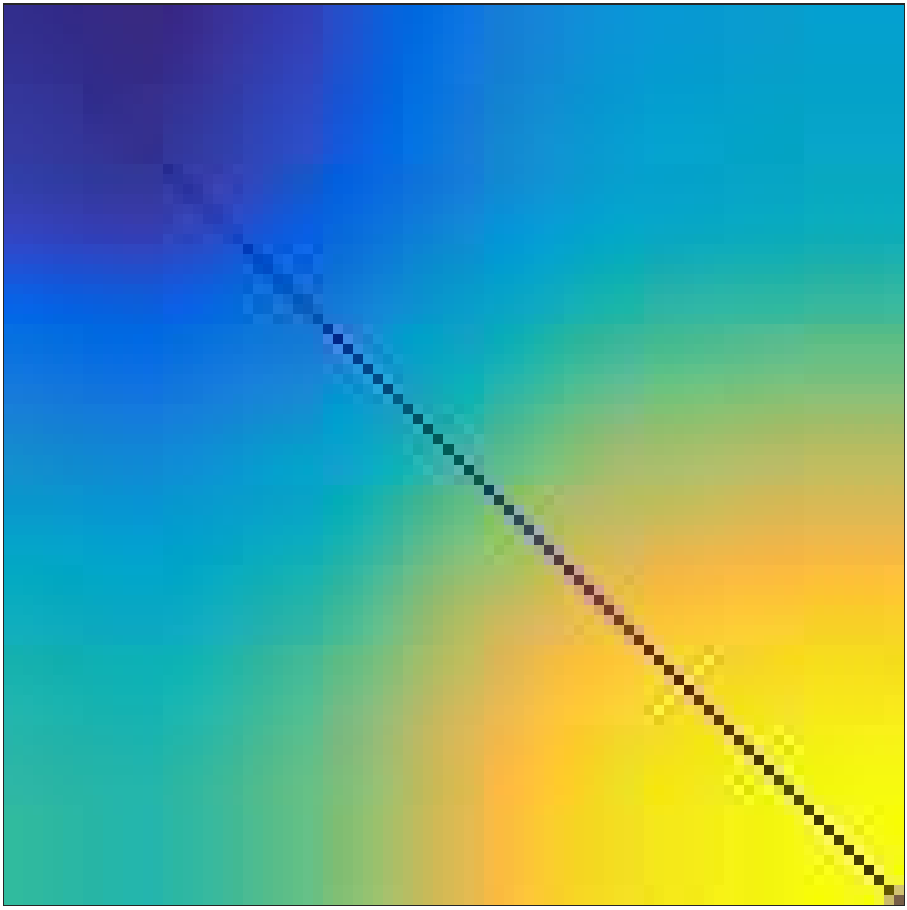}
	\end{tabular}
\caption{Correlation matrices (yellow means higher) between the predicted DT using \emph{CNN-DT-P(T)} and T using \emph{CNN-T} (left); and between the predicted DT using \emph{CNN-W-P(R)} and T using \emph{CNN-T} (right).}
\label{fig:transflearn4}
\end{figure} 



%% file: sections/references.tex
\small
\bibliographystyle{plain}
\bibliography{biblio}